Systematic Review
Irene S. Gabashvili
https://aurametrix.com

# The impact and applications of ChatGPT: a Systematic Review of Literature Reviews

## Abstract

**Background:** The conversational artificial-intelligence (AI) technology ChatGPT has become one of the most widely used natural language processing tools. With thousands of published papers demonstrating its applications across various industries and fields, ChatGPT has sparked significant interest in the research community. Reviews of primary data have also begun to emerge. An overview of the available evidence from multiple reviews and studies could provide further insights, minimize redundancy, and identify areas where further research is needed.
**Objective:** To evaluate the existing reviews and literature related to ChatGPT's applications and its potential impact on different fields by conducting a systematic review of reviews and bibliometric analysis of primary literature.
**Information sources:** PubMed, EuropePMC, Dimensions AI, medRxiv, bioRxiv, arXiv, and Google Scholar were searched using ChatGPT-related keywords from November 2022 to the end of April 2023.
**Eligibility criteria:** Studies including secondary data (such as systematic reviews, meta-analyses and bibliometric analyses) related to the application of ChatGPT. Articles published in all languages were considered.
**Methods:** Reporting was performed using PRISMA (Preferred Reporting Items for Systematic Reviews and Meta-Analyses) guidelines. Analysis of study findings was conducted using the thematic synthesis approach. The risk of bias was addressed by applying the PRISMA and AMSTAR (Assessing the Methodological Quality of Systematic Reviews) tools.
**Results:** A total of 305 unique records with potential relevance to the review were identified from a pool of over 2,000 original articles. After a multi-step screening process, we selected 11 reviews, including 9 specifically focused on ChatGPT (encompassing 80 to 385 unique records identified and 10 to 60 records subjected to systematic analysis, in addition to 186 records analyzed through bibliometric methods), along with 2 reviews on broader AI topics that also discussed ChatGPT. Additionally, we conducted bibliometric analysis on 1,357 ChatGPT papers out of a total of 1,854,007 records published from December 2022 to April 2023.
**Conclusions:** While AI has the potential to revolutionize various industries, further interdisciplinary research, customized integrations, and ethical innovation are necessary to address existing concerns and ensure its responsible use.
**Protocol Registration:** PROSPERO registration no. CRD42023417336,
Centre for Open Science DOI 10.17605/OSF.IO/87U6Q.

# Introduction

## Background

Artificial intelligence (AI) has shown a positive impact on various fields, and pre-trained language models have become the new paradigm of natural language processing (NLP). The basis for the new AI technology called "ChatGPT" is an advanced version of the generative pre-trained transformer (GPT) architecture, which has been found to possess remarkable emergent skills. These skills include the ability to follow directions, which enables the production of complex and tailored behavior for various applications. As a result, ChatGPT is considered superior to its predecessors in terms of its capabilities.

ChatGPT has generated mixed responses, reflecting the history of controversy regarding the benefits vs. risks of AI. Less than 6 months since its launch, over two thousand papers have been published or posted on preprint servers. Multiple reviews have already been conducted to explore the future perspectives of ChatGPT, identify potential limitations and concerns, and investigate its utility in various domains. These reviews highlight the potential limitations and concerns associated with the application of ChatGPT in various fields. Furthermore, they present the growing trend of using ChatGPT in various domains, including education, health, finance, and academic writing.

## Goals

The primary objective of this study is to evaluate the existing reviews and literature related to ChatGPT's applications and its potential impact on different fields. This will be accomplished by conducting a systematic review of reviews and a bibliometric analysis of primary literature. The results of these analyses will be used to better understand the potential impact of ChatGPT in various fields, identify the most promising future applications of ChatGPT and to identify areas where further research is needed.

Another aim of this meta-review is to investigate methods for processing scientific literature more efficiently. We plan to experiment with using ChatGPT for different stages of the review and synthesis process, with the goal of finding ways to expedite the analysis of large volumes of scientific literature.

# Methods

## Literature Search Strategy and Study Design

We performed a systematic review consistent with the reporting guidance for conducting a systematic review of reviews. This involved a comprehensive search of academic databases and grey literature to identify relevant reviews of ChatGPT's applications in different fields. The selected reviews were critically appraised, and data were extracted and synthesized using thematic analysis.

To retrieve articles published between November 2022 and 30 April 2023, several repositories were searched for relevant articles and reviews about ChatGPT. The databases included Medrxiv, Biorxiv, EuropePMC, Pubmed, Dimensions AI, and Google Scholar.

In arXiv, MedrXiv and BiorXiv, the search keyword used was "ChatGPT" since "Chat GPT" returned too many irrelevant results. In EuropePMC, the search query used was (chatgpt OR "Chat GPT"). For Pubmed, the search query used was ChatGPT OR (Chat GPT). The filters used for review extraction in Pubmed were Meta-Analysis, Review, and Systematic Review.

For Dimensions AI, the search was conducted for articles published in the years 2022 and 2023 and the search query used was ChatGPT OR (Chat GPT) in Title and Abstract. In Google Scholar, the search keyword used was ChatGPT and the review filter was used to extract potential review articles.

Since ArXiv, MedrXiv, BiorXiv, and Dimensions AI did not have review filters, initial screening of titles was done programmatically by looking at "review," "meta-analysis," and bibliometric keywords.

The protocol was registered for health-related outcomes on PROSPERO on April 15 2023 (registration DOI: CRD42023417336, available from: https://www.crd.york.ac.uk/prospero/display_record.php?ID=CRD42023417336 and for more broad applications on the Centre for Open Science, registered on May 5 2023 (DOI 10.17605/OSF.IO/87U6Q).

Final search was conducted on April 30, 2023.

### Inclusion and Exclusion Criteria

To be included, studies had to be papers reviewing ChatGPT. We did not exclude reviews that did not meet all of the PRISMA and AMSTAR criteria due to the short time since ChatGPT's launch. Including reviews of "relatively poor" reporting quality is important for a comprehensive depiction of the state of the art. Additionally, some reviews may adequately report the analysis and synthesis of normative information but may not meet certain basic PRISMA and AMSTAR criteria.

Studies published in all languages were considered for inclusion

## Study Selection

The selection process was carried out in a similar way to reviews of individual studies.

To reduce bias in review selection, the review team usually consists of more than one member. We worked with ChatGPT as another member of the team.

Titles were checked for relevance and duplication. Similar titles belonging to different versions of the same article were identified and preprint versions were removed. In cases of duplicate publication, only the most recent and complete data was included.

Based on the above-mentioned criteria of inclusion and exclusion, the human author screened the list of unique titles and selected abstracts and full text retrieved through electronic and manual searches. The titles were then screened by ChatGPT according to the same criteria. Any discrepancy was solved by repeated review/better prompt engineering and abstract processing in order to reach a consensus. As seen from the PRISMA flow diagram for systematic reviews (Figure 1), a total of 142 potentially relevant reviews were retrieved.

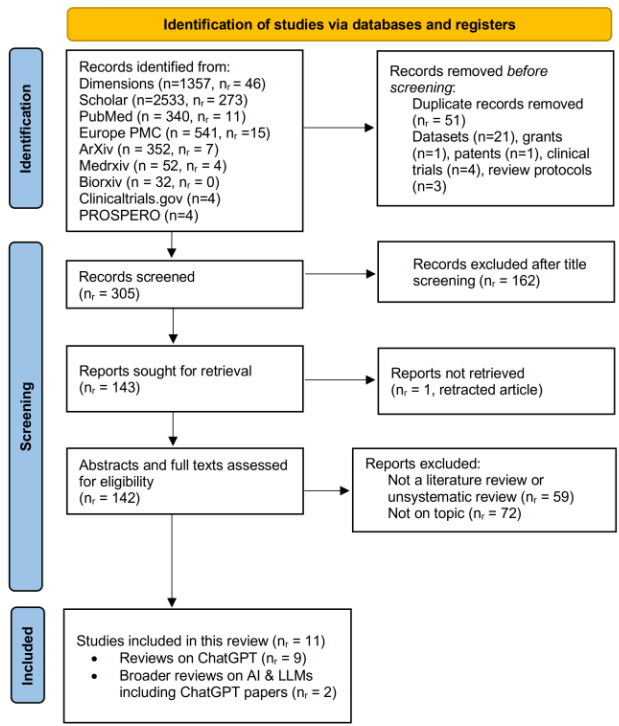

*Figure 1 PRISMA 2020 flow diagram for new systematic reviews. n is number of primary records, $n_r$ is number of reviews*

The further examination of the retrieved papers allowed us to exclude editorials, perspectives, unsystematic reviews and reviews not focusing on ChatGPT. Many of these papers mentioned ChatGPT in acknowledgements, methods or even as a coauthor. Some of the seemingly relevant reviews were focusing on broader topics including ChatGPT, but since their final literature search was performed before ChatGPT papers were available, their ChatGPT discussions were not supported by evidence. These reviews were also excluded as "Not on topic".

11 articles remained and were included in this review.

### Data Extraction & Synthesis

Full text of selected articles was carefully screened. Items such as author affiliations, countries, subject matter expertise, the time frame of included studies, type of measures, and theoretical background was collected included along with the details of literature search, data sources, and review outcomes.

Quality assessment questions included details of review selection and inclusion criteria, assessment of publication bias, discussion of heterogeneity tests, and comparability of included reviews in terms of eligibility criteria, study characteristics, and primary outcome of interest.

Thematic synthesis approach was used to combine key findings in the papers reviewed and identified key themes to explore in the intervention studies
Data were extracted by the human reviewer in standardized summary tables and were also categorized by ChatGPT. Any disagreements were re-reviewed.

Firstly, we carefully analyzed a sample of the studies we were reviewing, to understand their context and meaning. Then, we used this understanding to analyze the rest of the studies. The human reviewer coded all the data. ChatGPT was then asked to perform the same task. If there were any disagreements, the human reviewer reviewed the data again and reformulated the prompts until an agreement was reached. In the second phase, we identified common themes and topics in the studies, and grouped the codes into descriptive themes. We organized these themes into a table, which helped us understand the similarities and differences between the individual studies. In the third phase, we identified interpretive or analytical themes based on the insights we gained from the previous phases.

In addition to the review of reviews, we also conducted a bibliometric analysis of primary literature related to ChatGPT. This analysis involved identifying and analyzing the citation patterns, authorship, and publication trends in the primary literature. In particular, we compared the number of ChatGPT articles published in different Fields of Research to the total number of articles published in the same Fields of Research during the same time period from the Dimensions database.

### Results
Fig. 1 summarizes the study selection process. Of the 11 studies included, 9 were on ChatGPT [1-9] and 2 were broader reviews that discussed ChatGPT [10-11]. Although the majority of articles on ChatGPT were in the English language, language barriers are falling as we also reviewed primary studies in Chinese, German, Indonesian, Norwegian, Portuguese, Russian, and Spanish. Thanks to evolving translator software, language is no longer a limitation.

Of the eleven titles considered for inclusion in this synthesis, five had "systematic review" in their title, one was named a rapid review, but its quality was comparable to that of the systematic reviews, one was a hybrid review, one was a science mapping, and one was a meta-analysis. Two high quality papers did not mention "review" in their title; one elaborated that although the author attempted to conduct a systematic review, due to insufficient evidence, their article may be considered a viewpoint, while the other noted the intention to systematically review literature in the abstract. All reviews were submitted or published between February and April of 2023.

**Overall sentiment**

Table 1 presents a list of analyzed articles along with their overall sentiment ratings as evaluated by ChatGPT (found to possess impressive abilities in understanding the emotions and inferring the emotion causes in [12] and by the author of this paper). The last column contains the most representative sentence for overall sentiment taken from the respective article.

*Table 1 Overall sentiment of ChatGPT reviews*

| Reference | Overall sentiment | Representative sentence(s) |
| --- | --- | --- |
| [1] ChatGPT Utility in Healthcare Education, Research, and Practice | Slightly Positive. | Health care professionals could be described as carefully enthusiastic regarding the huge potential of ChatGPT among other LLMs in clinical decision-making and optimizing the clinical workflow |
| [2] ChatGPT in Scopus and WoS: till 02/2023 | Slightly Positive. | In the future ChatGPT could definitely improve or even replace the consulting field. |
| [3] ChatGPT in Healthcare | Slightly negative. | Healthcare researchers in particular should retract from the AI hype generated by the product and focus their attention on NLP research in general and developing/evaluating specialized language models for healthcare applications |
| [4] ChatGPT in WoS & Scopus: till 3/31/2023 | Positive. | The study found that although ChatGPT has some limitations and requires human assistance at times, it has the potential for various applications. |
| [5] Impact of ChatGPT in Education | Neutral. | The findings of this review suggest that ChatGPT's performance varied across subject domains, ranging from outstanding (e.g., economics) and satisfactory (e.g., |
| | | programming) to unsatisfactory (e.g., mathematics) |
| [6] Early research trends on ChatGPT: 3/8/2023 | Positive. | Most works in different clusters discussed apprehension and excitement related to ChatGPT's potential to revolutionize education, especially medical writing and education. |
| [7] ChatGPT in Supply Chains | Positive. | Although it may take time until this technology evolves to a desirable level of maturity, it may be applied in different areas of supply chain management. |
| [8] ChatGPT in Medical Literature | Neutral. | … growing body of literature on ChatGPT's applications and implications in healthcare highlighting the need for further research to assess its effectiveness and ethical concerns. |
| [9] Perception of ChatGPT | Neutral/Slightly Positive. | Papers in the categories "Application, Medical and Rest" mostly describe ChatGPT as an opportunity for society. For Education, the number of papers that see ChatGPT as a threat is almost equal to the number of those that view it as opportunity. For Evaluation, a comparably high number of abstracts articulate mixed sentiments towards the social impact. Finally, in the Ethics category, ChatGPT is mostly seen as a threat. |
| [10] AI in Healthcare and Medicine | Slightly negative. | While the progress of fundamental AI methods provides great opportunities for medical AI, there is a severe concern about the reliability, safety, and factuality of generated content by medical AI methods, and significant challenges still exist in investigating the factuality problem in medical AI |
| [11] LLMs in education | Positive. | LLMs-based innovations have already shown high performance on several relatively simple classification tasks that could potentially be deployed to automatically generate meaningful insights that could be useful to teachers and institutions, such as navigating through numerous student feedback and course review. |

Two months after the launch of ChatGPT, the first review of its publications was posted to arXiv [9]. The review provided a sentiment analysis of over 300,000 tweets and an analysis of ChatGPT's impact based on 152 scientific papers published at that time. The analysis found that the average sentiment of the tweets was slightly more positive than neutral, with the most positive sentiment observed for business, entrepreneurship, science, and technology. The majority of papers analyzed reported a robust performance of ChatGPT, highlighting its potential to have a positive social impact. This sentiment was frequently described as an opportunity in Semantic Scholar Publications. However, the sentiment was more balanced in Arxiv, with an equal number of preprints identifying ChatGPT as an opportunity and a threat. ChatGPT determined sentiment of the conclusion of this paper to be neutral, but changed it to "slightly positive" after reading results and discussions, aligning with the human author's annotation.

### Analytical themes

Common themes identified across included studies were fields of research (FoR), areas of promising applications, mitigation strategies to address the practical challenges in using ChatGPT, ethical challenges, types and quality of articles and geography of authors.

Fields of Research (FoR) Figure 2 displays results of bibliometric analysis performed in this paper: a treemap diagram of relative percentage of ChatGPT articles in different Fields of Research. The analysis is based on 22 general categories representing 1,357 papers with ChatGPT mentioned in their titles and abstracts, out of a total of 1,854,007 records published between December 2022 and April 2023.

11 reviews included in this paper are written by scientists with expertise in at least half of these fields. Their review papers are focusing on an even smaller number of fields, mostly in education (medical, financial, sciences, etc), health and research, including academic writing.

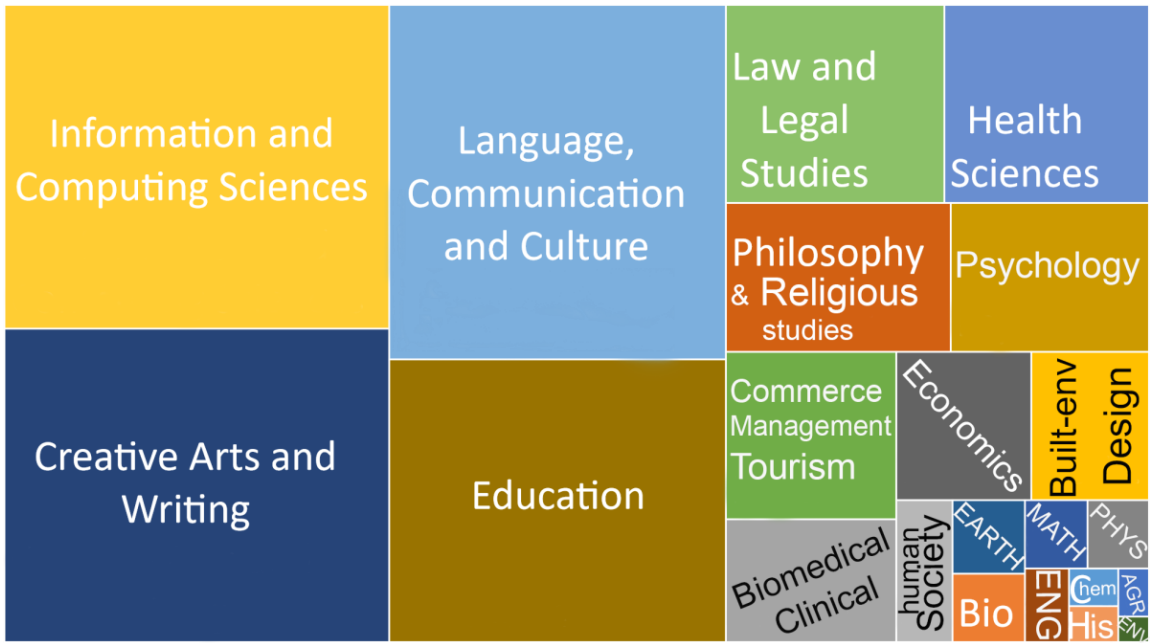

*Figure 2 Visual representation of ChatGPT's research impact across various fields of study, depicted through a treemap diagram showing the relative percentage of articles published in each field*

Bibliometric analysis of papers published before March 8, 2023 [6], determined the largest clusters of publications focused on biomedical, clinical, and health sciences, information and computing, language, communication, and culture, philosophy and religious studies. Across all these clusters most excitement related to ChatGPT's potential to revolutionize medical writing and education.

Table 2 summarizes application areas and suggestions on future research in the 11 reviews.

*Table 2 Date of final searches, unique records identified and selected for review, applications of ChatGPT and future directions of research.*

| Reference/Final Search Date | Records identified/included | Promising applications | Future directions |
|---|---|---|---|
| [1] 2/16/2023 | 280/60 | healthcare: education, research, and practice. | improve quality of training datasets, the ability of ChatGPT to provide justification for incorrect decisions |
| [2] 02/2023 | 80/35 | education, health, writing, finance and investments | customized integrations, domain-specific research |
| [3] 03/20/2023 | 140/58 | healthcare: consultation, research | More laboratory testing of different use scenarios, BioGPT, Contrastive Language-Image Pre-Training (CLIP), using application-oriented and user-oriented taxonomy for analysis of literature |
| [4] 3/31/2023 | 273/45 (15) | healthcare: personalized recommendations, answering medical questions, developing clinical decision and support systems, improving discharge summaries, helping with writing, translating, simulating interactions between employees and organizations, and supporting policy decision-making | interdisciplinary research at the intersection of artificial intelligence, digital technologies, and health |
| [5] 2/28/2023 | 363/50 | Education: tool for instructors, providing a starting point for creating course syllabi, teaching materials, and assessment tasks | use ChatGPT to generate raw materials to train course-specific chatbots; updating institutional policies |
| [6] 3/8/2023 | 385/183 | Medical writing and scientific publishing | ethics, mathematical problem-solving, critical thinking and analytical reasoning |
| [7] March 2023 | 138/10 | Supply chains: operations, communication, route optimization, predictive maintenance, invoicing, ordering process, customer service and workforce training | planning, sourcing, creating more sustainable and resilient supply chains during disruptive events, transition from Supply Chain 4.0 to Supply Chain 5.0 and ethical considerations in implementing ChatGPT in supply chains. |
| [8] March 2023 | 175 | medical education, scientific research, medical writing, and diagnostic decision-making | effectiveness & ethical implications of using ChatGPT across different disciplines |
| [9] 2/1/2023 | 152 (12) | Medical domain, various scientific fields | investigating more dimensions besides sentiment and emotion over longer stretches of time, impact on society, potential to exacerbate and mitigate existing inequalities and biases. |
| [10] March 2023 | 865/37 (4) | healthcare: medical Q&A, text summarization & simplification, report generation | systematic evaluation benchmark on more tasks, multilingual multi-modal LLMs, automatic factuality evaluation methods for factual correctness, prompt |

|  |  |  | learning, and reinforcement learning from human feedback |
|---|---|---|---|
| [11] 12/31/2022 | 663/118 (10) | education: semantic-based approaches for generating meaningful questions that are closely related to the source contents | Prediction and generation tasks, open-sourcing, human-centered approach, practical and ethical innovations |

Academic writing is a topic frequently discussed in ChatGPT literature. The interest in ChatGPT's potential for writing systematic reviews [13-17] is not surprising given the challenges and time commitment involved in conducting them.

Review of Pubmed articles from December 2022 to March 2023 [8] identified eight themes including (1) medical writing, (2) medical education, (3) diagnostic decision-making, (4) public health, (5) scientific research, (6) ethical considerations of ChatGPT use, (7) ChatGPT's potential to automate medical tasks, and (8) criticism of accuracy and bias. The criticism of the limitations of ChatGPT in solving mathematical problems and developing content requiring critical thinking was most frequently found in non-academic environments [6].

One of the papers examined in this study [7] concentrates on the potential use of ChatGPT in supply chains. The author decided against using the term "review" in the title because there were no other articles that specifically focused on this topic in the research databases. Instead, the author analyzed articles related to the logistics industry, retail, and other subjects that involved complex operations similar to those found in supply chains. Based on this analysis, the author concluded that ChatGPT has the potential to streamline supply chain operations by reducing waste and inefficiencies.

One of the significant contributions of the reviews was offering suggestions to address the limitations of ChatGPT. These suggestions included methods for injecting more knowledge effectively, developing reliable automatic evaluation metrics, and employing reinforcement learning with human feedback [10]. Additionally, proposed solutions involved implementing new regulations and incorporating improvements in the underlying language model through NLP innovations [3].

Based on their findings, review focusing on education [11] proposed three recommendations for future studies, including updating existing innovations with the latest state-of-the-art models, embracing the initiative of open-sourcing models/systems, and adopting a human-centered approach throughout the developmental process.

Although the majority of papers written on ChatGPT and getting most attention was from the United States and the United Kingdom [4,6], The 11 papers reviewed in this study were written by authors from 12 countries: Australia, Brazil, Germany, Hong Kong, India, Iran, Jordan, Malaysia, Romania, Saudi Arabia, the United Arab Emirates, and the United States. Authorship was almost uniformly distributed among these countries, with the highest number of affiliations from Germany and the United States.

## Discussion

### Principal Findings

The principal finding of this systematic review of reviews on ChatGPT is that it has the potential to revolutionize various fields, but its use should be embraced with caution due to potential limitations and challenges associated with its use. While there has been a substantial amount of published literature on ChatGPT, many papers were editorials, letters, notes or short communications not backed by adequate data, as noted in [1,2,3,12]. The abundance of low-quality repeated content is concerning, raising the question that current practice of science should be reformed.

To make ChatGPT more effective in different domains, future efforts should be spent on improving the quality of training datasets, developing customized integrations, and domain-specific research. Further interdisciplinary research is needed to investigate the effectiveness and ethical implications of using ChatGPT across different disciplines, to evaluate its factual correctness using automatic evaluation methods, and to investigate its impact on society beyond sentiment and emotion.

### Limitations
Despite its contributions, this study has several limitations that need to be acknowledged.

Firstly, the rapid pace at which articles are published in this field is a concern that was observed in this review of reviews. The included reviews and primary publications are predominantly based on GPT-3.5 series, and with the recent release of GPT-4 and integrations like Wolfram plugin, the presented findings may not represent the current state of the field.

Secondly, the heterogeneity of the included reviews in terms of their focus, scope, methodology, and quality may compromise the generalizability of the results. While some reviews focus on specific applications of ChatGPT, others provide a broader overview of the technology and its potential uses. Additionally, the quality of the included reviews varies, with some reviews having higher risk of bias or limited scope.

Thirdly, the inclusion of preprints that have not undergone peer review may affect the reliability and accuracy of the results.

Fourthly, this systematic review was conducted by a single author, which may limit the interpretability of the results.

Lastly, the rapid growth of literature on ChatGPT is another limitation. As this technology is still in its early stages, many studies have been conducted over a short period of time. This may lead to a lack of long-term data and a limited understanding of the technology's potential applications and limitations.

### Comparison with Prior Work

To the best of the authors' knowledge, this is the first systematic review of reviews on applications and implementation of ChatGPT

Comparing with prior work, this paper provides an overview of the existing literature on ChatGPT, identifies key themes and trends, and includes bibliometric analyses to demonstrate the expansion of fields benefiting from ChatGPT.

### Conclusions

While ChatGPT is one of the most emergent and promising technologies, many questions still remain. It is an enabler and a game-changer, but there exist privacy and ethical concerns, possibility of bias, potential legal liabilities, and concerns about validity. To ensure transparency, integrity, and scientific rigor, we need to optimize collaborative intelligence and continued training of LLMs.

In summary, ChatGPT has the potential to revolutionize different fields, but its use should be embraced with caution due to potential limitations and challenges associated with its use. More interdisciplinary research is needed to evaluate its effectiveness and ethical implications and to investigate its impact on society. We hope that this systematic review of reviews will contribute to the advancement of research on ChatGPT and guide future directions in this field.

## Acknowledgements
The author thanks ChatGPT for assistance.

## Funding
This research received no external funding.

## Data Availability Statement
All papers reviewed are properly cited, linked and freely available for download.

## Conflicts of Interest
None declared.

## Abbreviations

AI: Artificial Intelligence
AMSTAR (Assessing the Methodological Quality of Systematic Reviews) tools
FoR: Field of Research
GPT: generative pre-trained transformer
LLM: Large Language Model
PRISMA (Preferred Reporting Items for Systematic Reviews and Meta-Analyses)

## Appendix 1